\documentstyle[12pt,aps]{revtex}
\begin{document}

\title{
\hfill\parbox[t]{2in}{\rm\small\baselineskip 14pt 
CEBAF-TH-96-08 \\CMU-HEP94-35 \\DOE/ER/40682-88 }
\vskip 1.3cm 
Strange Hadronic Loops of the Proton: \\ A Quark Model Calculation\\~ \\~
}

\vskip 1.0cm
 
\author{Paul Geiger}
\address{Department of Physics, Carnegie Mellon University,
Pittsburgh, Pennsylvania 15213\\~ }

\vskip 0.5cm

\author{Nathan Isgur}
\address{CEBAF, 12000 Jefferson Avenue, Newport News, Virginia, 23606}
\maketitle

\vspace{1.5 cm}
%%%%%%%%%%%%%%%%%%%%%%%%%%%%%%%%%%%%%%%%%%%%%%%%%%%%%%%%%%%%%%%%%%%%
\begin{center}  {\bf Abstract}  \end{center}
%%%%%%%%%%%%%%%%%%%%%%%%%%%%%%%%%%%%%%%%%%%%%%%%%%%%%%%%%%%%%%%%%%%%
\vspace{.4 cm}
\begin{abstract}

    Nontrivial $q \bar q$ sea effects have their origin in the low-$Q^2$
dynamics of strong QCD. We present here a quark model calculation of the 
contribution of $s \bar s$ pairs arising from a {\it complete} set of OZI-allowed 
strong $Y^*K^*$
hadronic loops to the net spin of the proton, to its 
charge radius, and to its magnetic moment. The calculation is performed
in an ``unquenched quark model" which has been shown to preserve the spectroscopic
successes of the naive quark model and to respect the OZI rule. 
We speculate that an extension of the calculation to the 
nonstrange sea will show that most of the ``missing spin" of the 
proton is in  orbital angular momenta.

\end{abstract}
\newpage

%%%%%%%%%%%%%%%%%%%%%%%%%%%%%%%%%%%%%%%%%%%%%%%%%%%%%%%%%%%%%%%%%%
\section {Introduction}
%%%%%%%%%%%%%%%%%%%%%%%%%%%%%%%%%%%%%%%%%%%%%%%%%%%%%%%%%%%%%%%%%%

   While providing a good description of low-energy strong interaction 
phenomena, the constituent quark model appears to be 
inconsistent with many fundamental characteristics of QCD.
   Foremost among these inconsistencies is a  ``degree of freedom 
problem":
   the quark model declares that the low energy spectrum of QCD 
is built from the degrees of freedom of spin-$1 \over 2$ fermions 
confined to a $q \bar q$ or $qqq$ system.  
   Thus, for mesons the quark model predicts -~-~- and 
we seem to observe -~-~- a ``quarkonium"  spectrum.  
   In the baryons it predicts -~-~- and we seem to observe -~-~- the 
spectrum of two relative coordinates and three spin-$1 \over 2$
 degrees of 
freedom.
   
   These quark model degrees of freedom are to be contrasted with
the most naive interpretation of QCD which would 
lead us to expect a low energy spectrum exhibiting 36 quark and 
antiquark degrees of freedom (3 flavors $\times$ 2 spins $\times$ 3 
colors for particle and antiparticle), {\it and} 16 gluon degrees 
of freedom (2 spins $\times$ 8 colors).  
   Less naive pictures exist, but none evade the first major 
``degree of freedom problem" that the 
gluonic degrees of freedom appear to be missing from the low energy spectrum.  
   This issue, being one of the most critical in nonperturbative QCD, 
is being addressed by many theoretical and experimental 
programs.

   The second major ``degree of freedom problem", and the one
which we address here, 
has to do with $q \bar q$ pair creation.  
{\it A priori}, one would expect pair 
creation to be so probable that a valence quark model would 
fail dramatically.  Of course, we know empirically 
that pair creation is  suppressed:  the observed 
hadronic spectrum is dominated by narrow resonances,  
while the naive picture would predict resonances with 
widths $\Gamma$ comparable to their masses $m$. 

    It is now widely
appreciated that the narrow resonance approximation can
 be rationalized in QCD within the $1/N_c$ 
expansion~\cite{LargeNc}: in the limit $N_c \rightarrow \infty$,
meson widths (for example) are proportional to $N_c^{-1}$ 
while their masses are independent of $N_c$. 
The demonstration proceeds by showing that 
hadron two-point functions are dominated by graphs 
in which the valence quark lines propagate from their 
point of creation to their point of annihilation without 
additional quark loops.  The dominance of such a 
``quenched approximation" is, however, not sufficient 
to underwrite the valence quark model:  in the chiral limit, 
such Feynman graphs in general receive important 
contributions from not only forward quark propagation, 
but also from ``Z-graphs".  (A ``Z-graph" is one in which 
the interactions first produce a pair and then 
annihilate the antiparticle of the produced pair 
against the original propagating particle).  Cutting 
through a large $N_c$ two-point function at a fixed time 
therefore would in general reveal not only the valence 
quarks but also a large $q \bar q$ sea.  The large $N_c$ expansion 
also leaves unanswered a more quantitative question.  While 
hadronic widths $\Gamma$ {\it are} normally small compared to 
hadronic masses $m$, they are typically comparable to 
the mass spacings between states in the hadronic 
spectrum.  It is thus surprising that the spectroscopy of a valence quark model can 
survive ``unquenching".

   There is another puzzle of hadronic dynamics which 
is reminiscent of this one:  the success of the OZI 
rule~\cite{OZI}.  
   A generic OZI-violating amplitude $A_{OZI}$ can 
also be shown to  vanish like $1/N_c$. 
  However, there 
are several unsatisfactory features of this 
``solution" to the OZI mixing problem~\cite{LipkinOZI}.  
Consider $\omega$-$\phi$ 
mixing as an example.  This mixing receives a 
contribution from the virtual hadronic loop process 
$\omega \rightarrow K \bar K \rightarrow \phi$, both 
steps of which are OZI allowed, 
and each of which scales with $N_c$ like 
$\Gamma^{1/2} \sim N_c^{-1/2}$.
The large $N_c$ result that this OZI violating amplitude 
behaves like $N_c^{-1}$ is thus not peculiar to large $N_c$:  
it just arises from ``unitarity" in the sense that the real and 
imaginary parts of a generic hadronic loop diagram will 
have the same dependence on $N_c$. The usual interpretation of the OZI rule in this
case ~-~-~-~ that ``double hairpin graphs" are dramatically suppressed ~-~-~-~ is
untenable in the light of these OZI-allowed loop diagrams. They expose the 
deficiency of the large $N_c$ argument since $A_{OZI} \sim \Gamma 
<<m$ is {\it not} a good representation of the 
OZI rule.  (Continuing to use $\omega$-$\phi$ mixing as 
an example, we note that $m_\omega - m_\phi$ is numerically comparable to a 
typical hadronic width, so the large $N_c$ result would 
predict an $\omega$-$\phi$ mixing angle of order unity in 
contrast to the 
observed pattern of very weak mixing which implies 
that $A_{OZI} << \Gamma <<m$.)

In our recent papers~\cite{GIpotential,GIonOZI}, 
we have studied the unquenching 
of the quark model, addressing in particular 
the impact of $q \bar q$ pair creation on quark model spectroscopy 
and on the OZI rule. In this paper we extend our previous work to calculate 
some effects of the
strange quark content of the proton induced by strong $s \bar s$ pair creation.
Since, as will be described in the next section, our model preserves the
spectroscopic successes of the quark model and is consistent with the OZI rule,
it provides a legitimate framework for the study of the $q \bar q$ sea. We
focus here on the $s \bar s$ sea both because it allows us to avoid
complexities associated with antisymmetrization with respect to the valence
quarks in the nucleons, and because it has recently received considerable
experimental attention.

     Our goals for this calculation, though ambitious, are limited. In particular, we
will address here  the  effects 
on the
strange quark helicity $\Delta s$, the strangeness charge radius $R_s^2$, and
the strangeness magnetic moment $\mu_s$  of a complete sum over the OZI-allowed $s
\bar s$ loops which contribute to two-point functions 
({\it i.e.}, of processes that correspond at the hadronic level to $p \rightarrow 
Y^*K^* \buildrel j\over {\rightarrow}  Y^{*'}K^{*'} \rightarrow p$, where $Y^*$ and $K^*$ represent 
generic $S=-1$ baryons and $S=+1$ mesons, respectively,
and where $\buildrel j\over{\rightarrow}$ indicates the action of the appropriate
current). In contrast, 
we are unable to discuss the
effects of pure OZI-forbidden  processes ({\it i.e.}, ones that do not proceed
through strong OZI-allowed meson loops). These include processes in which the $s \bar s$ pair is
directly created or annihilated in a color singlet state ({\it e.g.}, 
$p \rightarrow p \phi_{J^{PC}} \buildrel j\over{\rightarrow}  p \phi'_{J'^{P'C'}} \rightarrow p$ 
and 
$p \buildrel j\over{\rightarrow}  p \phi_{J^{PC}} \rightarrow p$
or 
$p \rightarrow p \phi_{J^{PC}}  \buildrel j\over{\rightarrow}   p$,
where $\phi_{J^{PC}}$ is an $s \bar s$ meson with quantum numbers ${J^{PC}}$.
The latter two of these processes correspond to pure OZI-forbidden vector-meson-dominance-type
graphs.
As was the case in our earlier studies of OZI violation~\cite{GIonOZI},
all such disconnected ``double hairpin" diagrams are outside of the scope of our model: we focus on the
naively much larger OZI-allowed loop diagrams. We  also do not discuss here processes in
which strange baryon-meson loops are directly created by the  probing current.
While such ``contact graphs" would in general exist, we show below that
none are required to make the contributions of our strong
$Y^*K^*$ loop graphs to $\Delta s$, $R_s^2$, or $\mu_s$ gauge invariant.

%%%%%%%%%%%%%%%%%%%%%%%%%%%%%%%%%%%%%%%%%%%%%%%%%%%%%%%%%%%%%%%%%%
\section { Unquenching the Quark Model:  Background}
%%%%%%%%%%%%%%%%%%%%%%%%%%%%%%%%%%%%%%%%%%%%%%%%%%%%%%%%%%%%%%%%%%

  The Introduction describes three puzzles associated 
with the nature and importance of $q \bar q$ pairs in low 
energy hadron structure:
  
\vskip 0.5cm
 1) the origin of 
the apparent  valence structure of hadrons (since even 
as $N_c \rightarrow \infty$, Z-graphs would produce pairs unless 
the quarks were heavy),
\vskip 0.25cm
 2) the apparent absence of unitarity corrections to
naive quark model spectroscopy, despite one's expectation
of mass shifts $\Delta m \sim \Gamma$  (where $\Gamma$
is a typical hadronic width), and
\vskip 0.25cm
 3) the systematic suppression of OZI-violating 
amplitudes $A_{OZI}$ relative to one's expectation (from unitarity) 
that $A_{OZI} \sim \Gamma$.

\vskip 0.5cm
  
\noindent In this section we describe the solutions we see to 
these puzzles.  The resulting picture forms the context 
of the new work described in this paper.

\bigskip
%%%%%%%%%%%%%%%%%%%%%%%%%%%%%%%%%%%%%%%%%%%%%%%%%%%%%%%%%%%%%%%%%%%%%
\subsection{The Origin of the Valence Approximation} 
%%%%%%%%%%%%%%%%%%%%%%%%%%%%%%%%%%%%%%%%%%%%%%%%%%%%%%%%%%%%%%%%%%%%%

   As already mentioned, a weak form of the valence 
approximation seems to emerge from the large $N_c$ 
limit in the sense that diagrams in which only valence 
quark lines propagate through hadronic two-point 
functions dominate as $N_c \rightarrow \infty$.  
   This dominance does not seem to correspond to the usual 
valence approximation since the Z-graph pieces 
of such diagrams will produce a $q \bar q$ sea.

Consider, however, the Dirac equation for a single light 
quark interacting with a static color source (or a single 
light quark confined in a bag).  This equation represents 
the sum of a set of Feynman graphs which also include 
Z-graphs, but the effects of those graphs is captured 
in the lower components of the single particle Dirac 
spinor. {\it I.e.}, such Z-graphs correspond to relativistic 
corrections to the quark model.  That such corrections 
are important in the quark model has been known for a 
long time~\cite{gA}.  For us the important point is that while 
they have quantitative effects on quark model 
predictions ({\it e.g.}, they are commonly held to be 
responsible for much of the required reduction of 
the nonrelativistic quark model prediction  
that $g_A=5/3$ in neutron beta decay), they do not qualitatively 
change the single-particle nature of the spectrum of 
the quark of our example, nor would they 
qualitatively change the spectrum of $q \bar q$ or $qqq$ systems.
Note that this interpretation is consistent with the fact that
Z-graph-induced $q \bar q$ pairs do {\it not} correspond to the usual
partonic definition of the $q \bar q$ sea since Z-graphs
vanish in the infinite momentum frame. Thus the $q \bar q$ sea
of the parton model is also associated with the $q \bar q$
loops of unquenched QCD.

\bigskip
%%%%%%%%%%%%%%%%%%%%%%%%%%%%%%%%%%%%%%%%%%%%%%%%%%%%%%%%%%%%%%%%%%%%%
\subsection{  The $\Delta m << \Gamma$ Problem}
%%%%%%%%%%%%%%%%%%%%%%%%%%%%%%%%%%%%%%%%%%%%%%%%%%%%%%%%%%%%%%%%%%%%%

   Consider two resonances which are separated by a mass 
gap $\delta m$ in the narrow resonance approximation.  In general 
we would expect that departures from the narrow resonance 
approximation, which produce resonance widths $\Gamma$, 
ought also to produce shifts $\Delta m$ of order $\Gamma$.   
   Yet even though a typical hadronic mass spectrum is 
characterized by mass gaps $\delta m$ of order 500 MeV, and typical hadronic 
widths are of order 250 MeV, this does not seem to happen.

   We have proposed a simple resolution of this 
puzzle~\cite{GIpotential}.  
   In the  flux tube model of Ref.~\cite{IsgPat}, 
the quark potential model arises from 
an adiabatic approximation to the gluonic degrees of 
freedom embodied in the flux tube.  
   For example, the standard heavy $Q \bar Q$ quarkonium potential 
$V_{Q \bar Q} (r)$ is the ground state energy $E_0(r)$ of the gluonic 
degrees of freedom in the presence of the $Q \bar Q$ sources 
at separation $r$.  At short distances where perturbation 
theory applies, the effect of $N_f$ types of light $q \bar q$ pairs 
is (in lowest order) to shift the coefficient of the 
Coulombic potential from
$\alpha_s^{(0)}(Q^2)=\frac{12\pi}{33 {\it ln}(Q^2/\Lambda_0^2)}$ to
$\alpha_s^{(N_f)}(Q^2)=
\frac{12\pi}{(33-2N_f) {\it ln}(Q^2/\Lambda_{N_f}^2)}$.  
The net effect of such pairs 
is thus to produce a {\it new} effective short distance $Q\bar Q$ 
potential.

Similarly, when pairs bubble up in the flux tube ({\it i.e.}, 
when the flux tube breaks to create a $Q\bar q$ plus $q\bar Q$ system 
and then ``heals" back to $Q\bar Q$), their net effect is to 
cause a shift $\Delta E_{N_f}(r)$ in the ground state gluonic energy 
which in turn produces a new long-range effective $Q\bar Q$ 
potential~\cite{LevelCrossing}.  

In Ref.~\cite{GIpotential} we showed that the net long-distance 
effect of the bubbles is to create a 
new string tension $b_{_{N_f}}$ ({\it i.e.}, that the potential 
remains linear).  Since this string tension is to be 
associated with the observed string tension, after renormalization 
{\it pair creation has no effect on the long-distance structure 
of the quark model in the adiabatic approximation}.  Thus 
the net effect of mass shifts from pair creation
is much smaller than one would naively expect from 
the typical width $\Gamma$:  such shifts can only arise from 
nonadiabatic effects.  For heavy quarkonium, these shifts can in 
turn be associated with states which are strongly 
coupled to  nearby thresholds.

We should emphasize that it was necessary to sum over very 
large towers of $Q\bar q$ plus $q\bar Q$ intermediate states 
to see that the spectrum was only weakly perturbed (after 
unquenching and renormalization).
In particular, we found that no simple truncation of the 
set of meson loops can reproduce such results.

\bigskip
%%%%%%%%%%%%%%%%%%%%%%%%%%%%%%%%%%%%%%%%%%%%%%%%%%%%%%%%%%%%%%%%%%%%
\subsection{ The Survival of the OZI Rule}
%%%%%%%%%%%%%%%%%%%%%%%%%%%%%%%%%%%%%%%%%%%%%%%%%%%%%%%%%%%%%%%%%%%%

The Introduction illustrates, via the example of 
$\omega$-$\phi$ mixing through a $K\bar K$ loop, why unquenching 
the quark model endangers the naive quark model's agreement 
with the OZI rule.  In Refs.~\cite{GIonOZI} we showed
how this disaster is naturally averted in the flux tube 
model through a ``miraculous" set of cancellations 
between mesonic loop diagrams consisting of apparently 
unrelated sets of mesons ({\it e.g.}, the $K\bar K$, 
$K\bar K^*+K^*\bar K$, 
and $K^*\bar K^*$ loops tend to strongly cancel against 
loops containing a $K$ or $K^*$ plus one of the four strange 
mesons of the $L=1$ meson nonets).

   Of course the ``miracle" occurs for a good reason.  In the 
flux tube model, where pair creation occurs in the $^3P_0$ state, 
the overlapping double hairpin graphs which correspond to 
OZI-violating loop diagrams (see Fig. 1), cannot contribute 
in a closure-plus-spectator approximation 
since the $0^{++}$ quantum numbers of the produced 
(or annihilated) pair do not match those of the initial and 
final state for any established nonet.  
   Ref.~\cite{GIonOZI} demonstrates that this approximation 
gives zero OZI violation in all but the (still obscure) $0^{++}$ nonet, 
and shows that corrections to the closure-plus-spectator 
approximation are small, so that the observed hierarchy 
$A_{OZI} << \Gamma$ is reproduced.
  
     We emphasize once again that such cancellations require 
the summation of a very large set of meson loop diagrams 
with cancellations between apparently unrelated sets of 
intermediate states.
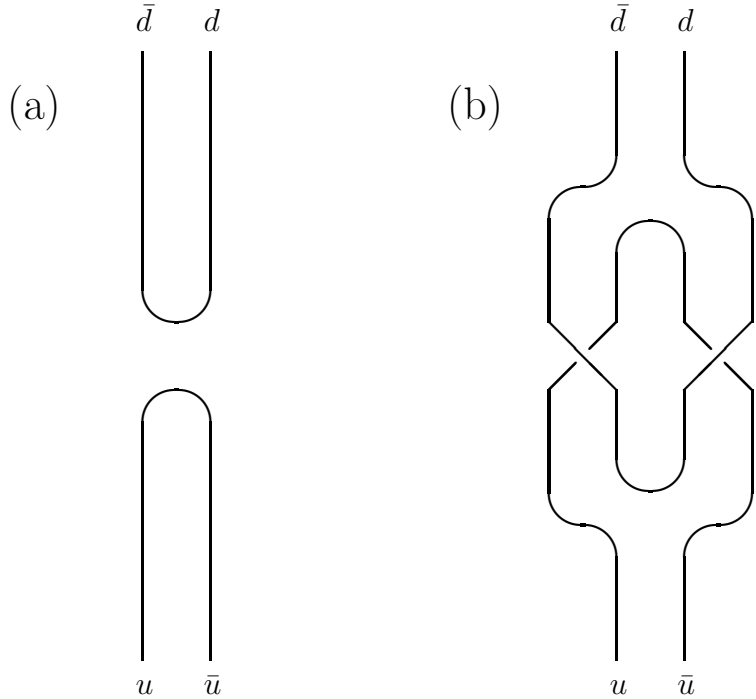
\begin{figure}
  \setlength{\unitlength}{0.9 mm}
  \begin{centering}
  \begin{picture}(120,130)(0,0)
    \thicklines
    \put(90,10){\oval(10,40)[tr]}  \put(110,10){\oval(10,40)[tl]} 
    \put(90,50){\oval(10,40)[bl]}  \put(110,50){\oval(10,40)[br]}
    \put(100,50){\oval(10,30)[bl]} \put(100,50){\oval(10,30)[br]} 
    \put(90,100){\oval(10,40)[br]} \put(110,100){\oval(10,40)[bl]} 
    \put(90,60){\oval(10,40)[tl]}  \put(110,60){\oval(10,40)[tr]}
    \put(100,60){\oval(10,30)[tl]} \put(100,60){\oval(10,30)[tr]} 
    \put(85,60){\line(1,-1){10}}   \put(115,60){\line(-1,-1){10}}
    \put(85,50){\line(1, 1){4}}    \put(95,60){\line(-1,-1){4}}
    \put(105,60){\line(1,-1){4}}   \put(115,50){\line(-1,1){4}}
    \put(94,5){$u$}                \put(104,5){$\bar{u}$}
    \put(94,103){$\bar{d}$}        \put(104,103){$d$}

    \put(30,10){\oval(10,80)[tl]}  \put(30,10){\oval(10,80)[tr]} 
    \put(30,100){\oval(10,80)[bl]} \put(30,100){\oval(10,80)[br]}
    \put(24,5){$u$}                \put(34,5){$\bar{u}$}
    \put(24,103){$\bar{d}$}        \put(34,103){$d$}
 
    \put(5,90){\Large (a)}         \put(70,90){\Large (b)} 
  \end{picture}
  \caption[x]{(a) OZI-violation in a meson propagator by ``pure
   annihilation'', corresponding to a disconnected double-hairpin diagram. 
(b) A different time ordering of the same
   Feynman graph gives an OZI-violating loop diagram via two OZI-allowed
   amplitudes. }
   \label{fig:uudd}
  \end{centering}
\end{figure}
\medskip
%%%%%%%%%%%%%%%%%%%%%%%%%%%%%%%%%%%%%%%%%%%%%%%%%%%%%%%%%%%%%%%%%%%%%
\subsection{Some Comments}
%%%%%%%%%%%%%%%%%%%%%%%%%%%%%%%%%%%%%%%%%%%%%%%%%%%%%%%%%%%%%%%%%%%%%

    We believe the preceding discussion strongly suggests 
that models which have not addressed the effects of unquenching on 
spectroscopy and the OZI rule should be viewed very 
skeptically as models of the effects of the 
$q \bar q$ sea on hadron structure:
we have shown that large towers of 
mesonic loops are required to understand how quarkonium 
spectroscopy and the OZI rule survive once strong pair 
creation is turned on.  In particular, while pion and kaon loops (which 
tend to break the closure approximation due to their 
exceptional masses) have a special role to play, 
they cannot be expected to provide a   
reliable guide to the physics of $q \bar q$  pairs.

\bigskip
%%%%%%%%%%%%%%%%%%%%%%%%%%%%%%%%%%%%%%%%%%%%%%%%%%%%%%%%%%%%%%%%%%
\section{  Strange Quarks and the Spin Crisis:  Some History}
%%%%%%%%%%%%%%%%%%%%%%%%%%%%%%%%%%%%%%%%%%%%%%%%%%%%%%%%%%%%%%%%%%

Beginning in 1988 with the European Muon Collaboration (EMC) 
experiment~\cite{EMC88}, and continuing through a recent 
series of closely related experiments~\cite{AfterEMC88},
the helicity structure functions of quarks in the 
proton and neutron have been measured via 
polarized deep inelastic scattering.
  When combined with measurements 
of  axial charges in hyperon beta-decay and the 
assumption of $SU(3)$ symmetry, these experiments 
indicate a ``spin crisis":  only about a third of 
the nucleon's helicity resides on its quarks, 
and about $ -10 \pm 3\%$ of this helicity is lost to 
strange quarks~\cite{EllKar}, in violation of the Ellis-Jaffe extension~\cite{EllJaf} of
the fundamental Bjorken Sum Rule~\cite{BjSum}.  

   Although generally accepted, there has been some discussion 
about the reliability of these conclusions.  
While support for them has come from other types of 
experiments~\cite{nuPexpt}, they have been criticized from other 
quarters for depending on an extrapolation of the 
structure functions to small $x$~\cite{CloseExtrap} and on an 
$SU(3)$-symmetry-based analysis of hyperon beta 
decay~\cite{LipkinSU3}. 
   At a deeper level, reanalyses of 
the theoretical connection between spin-dependent 
deep inelastic scattering and the spin structure 
functions showed that the $SU(3)$ singlet structure 
functions are entangled with the gluon spin structure 
functions via the $U(1)$ axial anomaly~\cite{anomaly}. This 
observation has
led to attempts to avert the ``spin crisis" by invoking 
a large gluonic contribution via the anomaly.  This 
possibility should be checked by direct measurements on 
the glue.

The naive nonrelativistic quark model predicts that 
$100\%$ of the nucleon's helicity resides on the quarks, 
but, as already mentioned above, lower components of 
the quark spinors arising from relativistic effects 
are believed to lower this fraction to about $75\%$~\cite{gA}.  
At the opposite extreme are naive Skyrmion models~\cite{EBK} which 
predict 
that the net quark spin of the nucleons should be zero 
(a result which seemed supported by the initial 
experimental results). 

If there is a large strange quark contribution to 
the nucleon spin, then one would also naturally expect
strange contributions to nucleon magnetic and electric 
form factors.  Purely electromagnetic scattering 
can only measure the four linear combinations
\begin{equation}
G^{\,\gamma p}_{M,E}= \frac{2}{3}G^{(u)}_{M,E}
-\frac{1}{3}G^{(d)}_{M,E}-\frac{1}{3}G^{(s)}_{M,E}
\end{equation}
\begin{equation}
G^{\,\gamma n}_{M,E}= \frac{2}{3}G^{(d)}_{M,E}
-\frac{1}{3}G^{(u)}_{M,E}-\frac{1}{3}G^{(s)}_{M,E},
\end{equation}
where $G^{(f)}_{M,E}$ is 
the magnetic ($M$) or electric ($E$) form factor of the 
quark flavor $f$ in the proton.  From parity violating 
scattering on the proton one can measure 
two more linear combinations
\begin{equation}
G^{Z p}_{M,E}=(\frac{1}{4}-\frac{2}{3}sin^2{\theta}_W )G^{(u)}_{M,E}
+(-\frac{1}{4}+\frac{1}{3}sin^2{\theta}_W )
[G^{(d)}_{M,E}+G^{(s)}_{M,E}]
\end{equation}
and thereby separate out the six elementary form 
factors $G^{(f)}_{M,E}$
for $f= u$, $d$, and $s$.  
Experiments are currently underway and others are 
planned to measure these form factors.  Such 
measurements appear to be the next step in 
understanding the physics of the spin crisis.

In the wake of the spin crisis have come a number 
of attempts to find theoretical descriptions less extreme than 
the naive quark and Skyrmion models.  In 1989, Jaffe~\cite{Jaffe1} 
pointed out 
that the  pole fit of  H\"ohler {\it et al.} 
to the nucleon's isoscalar 
electromagnetic form factors~\cite{Hohler} could suggest 
the presence of significant strange currents in the nucleon. 
  By identifying the two lightest fitted poles with the physical 
$\omega$ and $\phi$ mesons, he estimated 
$R_s^2 = 0.14\pm 0.07\;{\rm fm}^2$ for the strangeness 
radius and $\mu_s = -(0.31\pm 0.09) \;\mu_N$ for the strange 
magnetic moment of the nucleon. 
  (Note that Jaffe uses a sign convention in which the strange
quark has positive strangeness, opposite to the Particle Data Group's
convention~\cite{PDG}. As  a result, both $R_s^2$ and $\mu_s$
must be multiplied by $- {1 \over 3}$ to obtain the contributions
of the strange quarks to the electric and magnetic form factors
of the proton. Thus, for example, a positive 
value for $R^2_s$ indicates that $s$ quarks are farther 
on average from
the proton's center than $\bar{s}$ quarks. 
  Jaffe's convention appears to have been adopted by subsequent 
authors, and we too shall adhere to it.)

  Jaffe and Lipkin~\cite{JaffeLipkin}, building on
earlier work by Lipkin~\cite{Lipkin}, constructed an extended 
quark model in which the valence $qqq$ component of octet 
baryons was supplemented with a ``sea'' consisting of a single 
$q\bar{q}$ state which was allowed to have either $0^{++}$
or $1^{++}$ quantum numbers.
  Their model was not predictive; it was intended only as an 
example of a simple extension to the quark model which could 
accommodate the EMC results as well as baryon magnetic moments 
and hyperon $\beta$-decay.
  They found that the data could be fit with either 
a $(u\bar{u} +d\bar{d} +s\bar{s})$ or $(u\bar{u}+d\bar{d})$ 
flavour structure to the sea, though in both cases
a large suppression of the purely valence component
of the baryon wavefunctions was required. For other early
analyses along these lines, see Refs.~\cite{SiversCarlson}.

   The renormalization of axial couplings 
$g_A$ (and therefore of the fraction of the proton
spin $\Delta q$ carried by the quarks of flavor $q$) 
by $q \bar q$ pairs in the form of meson loops
is a subject with a
history dating back to the birth of meson exchange theories of the
strong interaction. For some modern studies in the context 
of chiral perturbation theory, see Ref.~\cite{JenMan}. Many 
recent studies, including ours, are extensions of this classic
meson loop approach~\cite{BjDrell}.  

   A model-dependent study of the $s \bar s$ sea based on
hyperon-kaon loop diagrams was made by Koepf {\it et al.} in 
Ref.~\cite{KHP}.
  These authors used both a non-relativistic quark model and the
cloudy bag model to calculate the strangeness content of the proton 
arising from $\Lambda K$, $\Sigma K$, and $\Sigma^* K$ loops.
  After tuning the baryon-baryon-meson form factors 
to reproduce the nonstrange nucleon moments, they found that
both models predict rather small strange moments:
$\Delta s \simeq -0.003$, 
$R_s^2 \simeq -.01 {\rm fm}^2$, and 
$\mu_s \simeq -.03 \mu_N $.

  Subsequently, Musolf and Burkhardt~\cite{MusBur} examined the
$\Lambda K$ loop graph in a calculation which took its vertex form
factors from the Bonn potential for baryon-baryon 
scattering~\cite{Bonn}, and which included seagull graphs in order
to satisfy the vector current Ward-Takahashi identity.
  The results obtained by these authors,
$\Delta s \simeq -0.044$,  
$R_s^2 \simeq -.03 {\rm fm}^2$, and $\mu_s \simeq -0.35
\mu_N$, are significantly larger than those 
found in~\cite{KHP}. The discrepancy is due, at least in part, to 
the non-gauge-invariance of the earlier calculation.
 
    More extensive, but from our perspective still incomplete, meson loop
calculations have been presented in Refs.~\cite{Speth}. These authors extend
loop calculations to include the entire set of ground state octet pseudoscalar
and vector mesons plus the ground state octet and decuplet baryons. They
obtain $\Delta s \simeq +0.02$. Another tack has been taken 
by Ito~\cite{Ito}, who calculates the effects of kaon loops on
a constituent quark, obtaining  
$R_s^2 \approx -.02\,{\rm fm}^2$ and $\mu_s \approx -.12
\,\mu_N$.
   
  The loop and pole pictures were combined 
in the model of Cohen {\it et al.}~\cite{Cohen}, who obtained
$R_s^2 \approx -.042\,{\rm fm}^2$ and $\mu_s \approx -.28\,\mu_N$.
These authors also calculated the strange form factors
at nonzero momentum transfer.

  For some reviews and for alternative models and points of view, 
in particular Skyrme-based calculations,
see Refs.~\cite{KitchenSink}.

\bigskip
\bigskip
%%%%%%%%%%%%%%%%%%%%%%%%%%%%%%%%%%%%%%%%%%%%%%%%%%%%%%%%%%%%%%%%%%
\section{ A Pair Creation Model for the Strangeness 
Content of the Proton}
%%%%%%%%%%%%%%%%%%%%%%%%%%%%%%%%%%%%%%%%%%%%%%%%%%%%%%%%%%%%%%%%%%

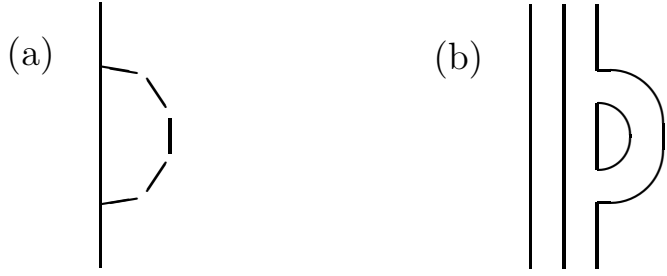
\begin{figure}
  \begin{centering}
  \setlength{\unitlength}{0.022cm}%
  \begin{picture}(404,220)(117,610)
  \thicklines
  \put(480,719){\oval( 80, 80)[br]}
  \put(480,719){\oval( 80, 80)[tr]}
  \put(480,719){\oval( 40, 40)[br]}
  \put(480,719){\oval( 40, 40)[tr]}
  \put(480,799){\line( 0,-1){ 40}}
  \put(480,739){\line( 0,-1){ 40}}
  \put(480,679){\line( 0,-1){ 40}}
  \put(460,799){\line( 0,-1){160}}
  \put(440,799){\line( 0,-1){160}}
  \put(208,754){\line( 2,-3){ 11.539}}
  \put(208,686){\line( 2, 3){ 11.539}}
  \put(181,761){\line( 6,-1){ 20.919}}
  \put(181,678){\line( 6, 1){ 20.919}}
  \put(222,730){\line( 0,-1){ 21}}
  \put(180,800){\line( 0,-1){160}}
  \put(117,762){$$ \large (a) $$}
  \put(375,760){$$ \large (b) $$}  
  \end{picture}
  \caption[x]{
  A meson loop correction to a baryon propagator,
  drawn at (a) the hadronic level, and (b) the quark level.}
  \label{fig:TheLoop}
  \end{centering}
\end{figure}

  Our discussion of the  strangeness content of the proton 
will be based on
the quark-level process shown in Fig.~\ref{fig:TheLoop}(b).
  The main new feature of our calculation is that we shall
sum over a {\it complete set} of strange intermediate states,
rather than just a few low-lying states. Not only does
this have a significant impact on the numerical results for 
$\Delta s$, $R^2_s$, and $\mu_s$, but, as explained above, 
it is {\it necessary} for consistency 
with the OZI rule and the
success of quark model spectroscopy.

\vspace{0.3cm}

  The lower vertex in Fig.~\ref{fig:TheLoop}(b) arises when
$q\bar{q}$ pair creation perturbs the initial nucleon state 
vector so that, to leading order in pair creation,
\begin{equation}
  |p\rangle \rightarrow |p\rangle +
  \sum_{Y^*K^* \ell S} \int q^2 dq\; 
  \left|Y^*K^* q\ell S \right\rangle 
  {\left\langle Y^*K^* q\ell S  \left| h_{q \bar q} \right| p \right\rangle
   \over M_p - E_{Y^*} - E_{K^*} } \; ,
  \label{eq:StateVec}
\end{equation} 
where $h_{q \bar q}$ is a quark pair creation 
operator, $Y^*$ ($K^*$) is the intermediate baryon (meson), $q$ and
$\ell$ are the relative radial momentum and orbital angular momentum of $Y^*$ and $K^*$,
and $S$ is the sum of their spins.
  Of particular interest is $s\bar{s}$ pair creation by the pair creation 
operator $h_{s \bar s}$, 
which will generate non-zero expectation values for strangeness 
observables:
\bigskip\bigskip
\begin{equation}
  \langle O_s \rangle = 
  \sum_{Y^*K^* \ell S \atop Y^{*\prime}K^{*\prime}\ell^{\prime}S^{\prime}} 
  \int q^2 dq\; {q^{\prime}}^2 dq^{\prime} \;
  {\left\langle p \left| h_{s \bar s} \right| 
  Y^{*^{\prime}}K^{*^{\prime}}q^{\prime} \ell^{\prime}S^{\prime}  \right\rangle
  \over M_p - E_{Y^{*^{\prime}}} - E_{K^{*^{\prime}}} }
  {\left\langle Y^{*^{\prime}}K^{*^{\prime}}q^{\prime}\ell^{\prime}S^{\prime} 
  \left| O_s \right| Y^*K^* q\ell S   \right\rangle}
  {\left\langle Y^*K^* q\ell S   \left| h_{s \bar s} \right| p 
  \right\rangle
   \over M_p - E_{Y^*} - E_{K^*} }.
  \label{eq:Os}
\end{equation} 
The derivation of this simple equation, including the
demonstration that it is gauge invariant, is given in the Appendix.
We will be considering the cases $O_s= \Delta s$, $R_s^2$, 
and $\mu_s$. The value of $\Delta s$ can be associated (via small scale-dependent
QCD radiative corrections) with the contribution of
strange quarks to the deep inelastic spin-dependent structure 
functions and to the strange quark axial current matrix elements
in the proton.

  To calculate the $p \rightarrow Y^*K^*$ vertices in 
Eq.~(\ref{eq:StateVec}), we employ the same flux-tube-breaking model
used in our earlier work.
  This model, which reduces to the well-known $^3P_0$ decay model 
in a well-defined limit, had its origin in applications
to decays of mesons~\cite{3P0mesons,KI} and 
baryons~\cite{3P0baryons}.
  The model assumes that a meson or baryon decays
when a chromoelectric flux tube breaks, creating 
a constituent quark and antiquark on the newly exposed flux
tube ends. The pair creation operator
is taken to have $^3P_0$ quantum numbers:
\begin{equation}
  h_{q \bar q}(t,{\bf x}) = \gamma_0 \left(3\over8\pi r_q^2\right)^{3/2}
  \int d^3z \; \exp\left(-{3z^2\over8r_q^2}\right) \;
  q^{\dagger}(t,{\bf x}+\frac{\bf z}{2}) 
  {\bf\alpha\cdot\nabla}
  q(t,{\bf x}-\frac{\bf z}{2}) \; . 
  \label{eq:Hpc}
\end{equation}
The dimensionless constant $\gamma_0$ is the intrinsic pair 
creation strength, a parameter which must be fit to 
decay data. In our previous studies of mesons, we fit 
$\gamma_0$ to the $\rho \rightarrow \pi\pi$ decay width;
here we find it more appropriate to fit to 
the $\Delta\rightarrow N\pi$ width. It turns
out that the two values agree to within 20\%, which is a
reassuring consistency check.  
  The operator (\ref{eq:Hpc}) creates {\it constituent} quarks, hence
the pair creation point is smeared out by a gaussian factor
whose width, $r_q$, is another parameter of the model.
  (In addition to being physically motivated, this smearing factor
is required  to render the sum in Eq.~(\ref{eq:Os}) finite.)
  As discussed in~\cite{GIpotential,GIonOZI}, $r_q$ is 
constrained  by meson
decay data to be approximately 0.25~fm.

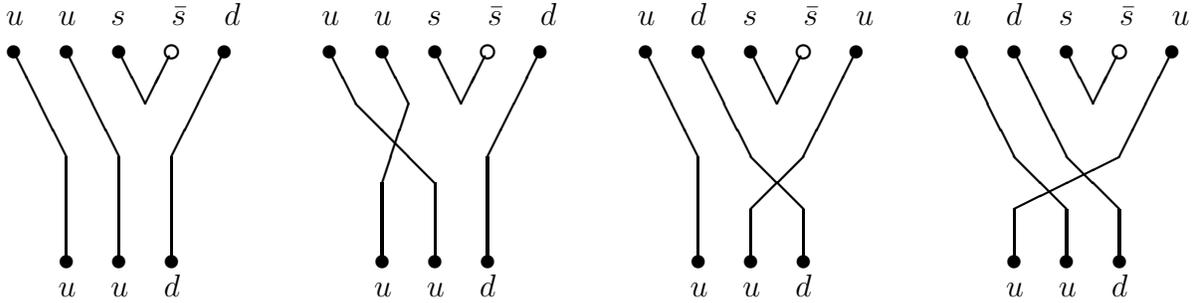
\begin{figure}
  \setlength{\unitlength}{0.35 mm}
  \begin{picture}(450,140)(30,0)
    \thicklines
    \put( 50, 20){\line( 0,1){40}}    \put( 50, 60){\line(-1,2){20}} 
    \put( 70, 20){\line( 0,1){40}}    \put( 70, 60){\line(-1,2){20}} 
    \put( 90, 20){\line( 0,1){40}}    \put( 90, 60){\line( 1,2){20}} 
    \put( 80, 80){\line(-1,2){10}}    \put( 80, 80){\line( 1,2){9}} 
    \put( 50, 20){\circle*{5}}        \put( 70, 20){\circle*{5}}
    \put( 90, 20){\circle*{5}}        \put( 30,100){\circle*{5}}
    \put( 50,100){\circle*{5}}        \put( 70,100){\circle*{5}}
    \put( 90,100){\circle{5}}         \put(110,100){\circle*{5}}
    \put( 47,  7){$u$} \put( 67,  7){$u$} \put( 87,  7){$d$}
    \put( 27,110){$u$} \put( 47,110){$u$} \put( 67,110){$s$}
    \put( 90,110){$\bar{s}$} \put(110,110){$d$}

    \put(170,20){\line( 0,1){30}}     \put(170, 50){\line( 1,3){10}}
    \put(180,80){\line(-1,2){10}}  
    \put(190,20){\line( 0,1){30}}     \put(190, 50){\line(-1,1){30}} 
    \put(160,80){\line(-1,2){10}} 
    \put(210,20){\line( 0,1){40}}     \put(210, 60){\line( 1,2){20}} 
    \put(200,80){\line(-1,2){10}}     \put(200, 80){\line( 1,2){ 9}} 
    \put(170, 20){\circle*{5}}        \put(190, 20){\circle*{5}}
    \put(210, 20){\circle*{5}}        \put(150,100){\circle*{5}}
    \put(170,100){\circle*{5}}        \put(190,100){\circle*{5}}
    \put(210,100){\circle{5}}         \put(230,100){\circle*{5}}
    \put(167,  7){$u$} \put(187,  7){$u$} \put(207,  7){$d$}
    \put(147,110){$u$} \put(167,110){$u$} \put(187,110){$s$}
    \put(210,110){$\bar{s}$} \put(230,110){$d$}

    \put(290,20){\line( 0,1){40}}     \put(290, 60){\line(-1,2){20}} 
    \put(310,20){\line( 0,1){20}}     \put(310, 40){\line( 1,1){20}} 
    \put(330,60){\line( 1,2){20}}  
    \put(330,20){\line( 0,1){20}}     \put(330, 40){\line(-1,1){20}} 
    \put(310,60){\line(-1,2){20}} 
    \put(320,80){\line(-1,2){10}}     \put(320, 80){\line( 1,2){ 9}}
    \put(290, 20){\circle*{5}}        \put(310, 20){\circle*{5}}
    \put(330, 20){\circle*{5}}        \put(270,100){\circle*{5}}
    \put(290,100){\circle*{5}}        \put(310,100){\circle*{5}}
    \put(330,100){\circle{5}}         \put(350,100){\circle*{5}}
    \put(287,  7){$u$} \put(307,  7){$u$} \put(327,  7){$d$}
    \put(267,110){$u$} \put(287,110){$d$} \put(307,110){$s$}
    \put(330,110){$\bar{s}$} \put(350,110){$u$}

    \put(410,20){\line( 0,1){20}}     \put(410, 40){\line( 2,1){40}} 
    \put(450,60){\line( 1,2){20}}
    \put(430,20){\line( 0,1){20}}     \put(430, 40){\line(-1,1){20}}
    \put(410,60){\line(-1,2){20}} 
    \put(450,20){\line( 0,1){20}}     \put(450, 40){\line(-1,1){20}}
    \put(430,60){\line(-1,2){20}} 
    \put(440,80){\line(-1,2){10}}     \put(440, 80){\line( 1,2){ 9}}
    \put(410, 20){\circle*{5}}        \put(430, 20){\circle*{5}}
    \put(450, 20){\circle*{5}}        \put(390,100){\circle*{5}}
    \put(410,100){\circle*{5}}        \put(430,100){\circle*{5}}
    \put(450,100){\circle{5}}         \put(470,100){\circle*{5}}
    \put(407,  7){$u$} \put(427,  7){$u$} \put(447,  7){$d$}
    \put(387,110){$u$} \put(407,110){$d$} \put(427,110){$s$}
    \put(450,110){$\bar{s}$} \put(470,110){$u$}

  \end{picture}
  \caption[x]{Quark line diagrams for $p\rightarrow\Sigma^* K^*$
  and $p\rightarrow\Lambda^* K^*$.}
  \label{fig:qlds}
\end{figure}

  Once an $s\bar{s}$ pair is created, the decay proceeds by 
quark rearrangement, as shown in 
Fig.~\ref{fig:qlds}. The $p\rightarrow Y^*K^*$ decay amplitude 
of the first of Figs.~\ref{fig:qlds} may be written as
\begin{equation}
  \langle Y^*K^*|h_{s \bar s}|p\rangle = \gamma_0\;\vec{\Sigma}\cdot\vec{I},
  \label{eq:ABC}
\end{equation}
where $\vec{\Sigma}$ is a spin
overlap which can be expressed in terms of the baryon and meson spin 
wavefunctions as
\begin{equation}
  \vec{\Sigma} \equiv \sum_{s_1 \cdots s_5} \; 
  \chi^{*Y^*}_{s_1 s_2 s_4} \;
  \chi^{*K^*}_{s_3 s_5} \; \chi^{p}_{s_1 s_2 s_3} \;
  \vec{\chi}_{s_4 s_5} \; ,
  \label{eq:spinsum}
\end{equation}
with
\begin{equation}
  \vec{\chi}_{s_4 s_5}  \equiv \left(
  \begin{array}{c}
   2\delta_{s_4{}_\uparrow} \delta_{s_5{}_\uparrow} \\
   -\delta_{s_4{}_\downarrow}\delta_{s_5{}_\uparrow}
   -\delta_{s_4{}_\uparrow}\delta_{s_5{}_\downarrow}\\
  -2\delta_{s_4{}_\downarrow} \delta_{s_5{}_\downarrow}
  \end{array}
  \right) \; , 
\end{equation}
and $\vec{I}$ a spatial overlap:
\begin{eqnarray}
  \vec{I} &=& 2\gamma_0 \left({3\over 4\pi b}\right)^{3/2}
  \int d^3k \, d^3p \, d^3s \; 
  \exp\left({-s^2\over 2b}\right) \; 
  \Phi^*_{Y^*}\left[{\bf k}, \; -\sqrt{3\over 2} 
  \left({\bf p}-{{\bf s}\over 2} - {m_s\over m_{uus}}{\bf q}
  \right)\right] \; 
  \nonumber \\ \nonumber \\
  && 
  \times \Phi^*_{K^*}\left[{\bf p} +{{\bf s}\over 2} - 
    {m_s\over m_{us}}q \right] \;
  {\bf p} \exp\left( -{2\over 3} r_q^2 p^2 \right) \;
  \Phi_p\left[ {\bf k},\; -\sqrt{3\over 2} 
    \left({\bf p}+{{\bf s}\over 6} -{\bf q} \right)\right] .
  \label{eq:vecI} 
\end{eqnarray}
Here the $\Phi$'s are momentum space wavefunctions, 
{$\bf q$} is the momentum of $Y^*$, and the $m_i$'s are
quark masses ($m_{uus}$ is short for $2m_u + m_s$, {\it etc.}).
  The factor $\exp(-s^2/2b)$ models the overlap of the initial and
final-state flux tube wavefunctions; its size is controlled by
the physical string tension $b$, though our
results do not depend strongly on the numerical value of $b$.

  For the remaining quark line diagrams in Fig.~\ref{fig:qlds}, 
the decay amplitude still has the form (\ref{eq:ABC}), but
the spin indices in Eq.~(\ref{eq:spinsum}) become permuted.
(The spatial overlap in (\ref{eq:vecI}) remains the same thanks to
the assumed symmetry of the proton's spatial wavefunction.)

\vspace{0.5cm}
 Faced with the large number of states that contribute to the sum
in Eq.~(\ref{eq:Os}), we have found it necessary to 
use simple harmonic oscillator (SHO) wavefunctions for the baryons
and mesons in (\ref{eq:vecI}). The  oscillator parameters
$\beta$ (defined by $\Phi({\bf k})\sim e^{-k^2/2\beta^2}$),
were taken to be $\beta_{meson}=0.4$ GeV for mesons (as in Ref.~\cite{KI})
and $\beta_{baryon}=0.32$ GeV for baryons (as in Ref.~\cite{beta32}). 
As discussed below,
our results are quite insensitive to changes in the $\beta$'s
(mainly because Eq.~(\ref{eq:Os}) is independent of the choice
of wavefunctions in the closure limit -~-~- any complete set gives
the same result -~-~- and the full calculation with energy denominators
does not deviate much from this limit.)
   
   Even with SHO wavefunctions, the sum over intermediate states would
be very difficult were it not for an important selection rule: inspection
of the quark line diagrams in Fig.~\ref{fig:qlds} shows that the 
relative coordinate of the non-strange quarks in baryon $Y^*K^*$ is always
in its ground state. Only the relative coordinate between the strange
and non-strange quarks ({\it i.e.}, the $\lambda_{Y^*}$-oscillator) can
become excited. This drastically reduces the number of states that
must be summed over. Unfortunately, this simplification
does not apply for
$u\bar{u}$ or $d\bar{d}$ pair creation; we therefore
postpone to a later paper the computation of their 
contributions to the nucleons' spin, charge radii, 
and magnetic moments.

   We will find it useful at times to refer to the
closure-spectator limit of Eq.~(\ref{eq:Os}). This is the limit in
which the energy denominators do not depend
strongly on the quantum numbers of $Y^*$ and $K^*$, so that 
the sums over intermediate states collapse to 1, giving
\begin{equation}
  \langle O_s \rangle \propto
  \left\langle p \left| h_{s\bar{s}} O_s 
   h_{s\bar{s}} \right| p\right\rangle \propto \left\langle 0 \left| h_{s\bar{s}} O_s 
   h_{s\bar{s}} \right| 0 \right\rangle \; ,
  \label{eq:OsClosure}
\end{equation}
where the second step follows since $h_{s\bar{s}}$ does not couple to the motion of
the valence spectator quarks.
We see that the expectation value of $O_s$ is taken
between the $^3P_0$ states created by $h_{s\bar{s}}$.
  From the properties of the $^3P_0$ wavefunction it then follows 
that $\Delta s = R_s^2 = \mu_s = 0$ in the closure-spectator limit (a result 
which would not be seen if only the lowest term, or lowest few 
terms, were included in the closure sum).

   In the next Section we will present our results for the 
expectation values defined by Eq.~(\ref{eq:Os}) for the quantities 
$\Delta s$, $R_s^2$, and $\mu_s$. We will see that delicate cancellations
lead to small values for these observables
even though the probability of $s \bar s$
pairs in the proton is of order unity!

%%%%%%%%%%%%%%%%%%%%%%%%%%%%%%%%%%%%%%%%%%%%%%%%%%%%%%%%%%%%%%%%%%
\section{Results}
%%%%%%%%%%%%%%%%%%%%%%%%%%%%%%%%%%%%%%%%%%%%%%%%%%%%%%%%%%%%%%%%%%
\label{sec:Results}

\subsection{Strange spin content}
  $\Delta s$, the fraction of the proton's spin carried by strange
quarks, is given by twice the expectation value of the $s$ and 
$\bar{s}$ spins :
\begin{equation}
 \Delta s = 2\left\langle 
            S_z^{(s)} + S_z^{(\bar{s})} 
            \right\rangle .
  \label{eq:Deltas}
\end{equation}
   Let us first examine the contribution to $\Delta s$ from 
just the lowest-lying intermediate state, 
$\Lambda K$. 
The $P$-wave 
$\Lambda K$ state with $J =J_z ={1\over 2}$ is
\begin{equation}
  \left| (\Lambda K)_{P {1\over2}}\right\rangle  = 
  \sqrt{2\over3}\Bigl|(\Lambda_{\downarrow}K)_{m=1}
  \Bigr\rangle 
 -\sqrt{1\over3}\Bigl|(\Lambda_{\uparrow}K)_{m=0}
  \Bigr\rangle .
  \label{eq:LamK}
\end{equation}
The $\bar{s}$ quark in the kaon is unpolarized, while the $s$
quark in the $\Lambda$ carries all of the $\Lambda$'s spin; 
because of the larger coefficient multiplying the first term in 
(\ref{eq:LamK}), the $\Lambda K$ intermediate state alone 
gives a negative contribution to $\Delta s$.
  
  When we add in the $(\Lambda K^*)_{P {1\over2}}$ and
$(\Lambda K^*)_{P {3\over2}}$ states (note that 
the subscripts denote the quantities $\ell S$ defined previously), we have 
\begin{equation}
   \Delta s \; \propto \;
   \left( \matrix{1 & -\sqrt{1\over 3} & \sqrt{8\over 3} \cr} 
   \right)\;\;
   {1\over 18} 
   \left[ \matrix{ -3 & \sqrt{3} & -\sqrt{24} \cr
                      &    -1    &  \sqrt{8}  \cr
                      &          &     10     \cr
                      }\right]
   \left( \matrix{1 \cr -\sqrt{1\over3} \cr \sqrt{8\over3} \cr} 
   \right)
   \label{eq:DsFromLK}
\end{equation}
in the closure limit. Here the matrix is just 
$2( S_z^{(s)} + S_z^{(\bar{s})} )$ (which is of course symmetric
though we only show its upper triangle), 
and the vectors give the relative 
coupling strengths of the proton to 
$[\; (\Lambda K)_{P {1\over2}}$, $(\Lambda K^*)_{P {1\over2}}$,
$(\Lambda K^*)_{P {3\over2}} \;]$. 
There are a couple of things to note here: 

  (1) The matrix multiplication in (\ref{eq:DsFromLK}) evaluates 
to zero; there is no net contribution to $\Delta s$ from 
the $\Lambda K$ and $\Lambda K^*$ states in the closure limit.
   There are in fact many such ``sub-cancellations'' in the
closure sum for $\Delta s$: for each fixed set of spatial quantum 
numbers in the intermediate state, the sum over quark spins alone 
gives zero (because 
$\langle S_z^{(s)} \rangle = \langle S_z^{(\bar{s})} \rangle = 0$
in the $^3P_0$ state).
  That is, each $SU(6)$ multiplet inserted into Eq.~(\ref{eq:Os})
separately sums to zero.
  Moreover, the $\Delta s$ operator does not cause transitions
between $I=0$ and $I=1$ strange baryons so that the $\Lambda$ and
$\Sigma$ sectors are decoupled, hence they individually sum 
to zero.

   (2) Only the diagonal term in Eq.~(\ref{eq:DsFromLK}) 
corresponding to 
$p \rightarrow (\Lambda K^*)_{P {3\over2}} {\buildrel \Delta s \over \rightarrow} 
(\Lambda K^*)_{P {3\over2}} \rightarrow p$
gives a positive contribution to $\Delta s$. (We use ${\buildrel \Delta s \over \rightarrow}$
here to denote the action of the $\Delta s$ operator.) All of the other terms
give negative contributions. In the full calculation with energy 
denominators, the negative terms are enhanced because they contain 
kaon (rather than $K^*$) masses. The full calculation gives
$\Delta s = -0.065$ from $\Lambda K$ and $\Lambda K^*$ states.
The largest individual contribution is $-0.086$, from the 
off-diagonal term  $p \rightarrow (\Lambda K)_{P {1\over2}} 
{\buildrel \Delta s \over \rightarrow} (\Lambda K^*)_{P {3\over2}} \rightarrow p$.

   Proceeding to intermediate states containing $\Sigma$ and 
$\Sigma^*$ baryons, we calculate 
\begin{equation}
  2( S_z^{(s)} + S_z^{(\bar{s})} ) = 
  {1\over 54} 
  \left[ \matrix{
  3 & -12\sqrt{2} & 3\sqrt{3} & -6\sqrt{6}&   0       &    0       \cr
    &     15      &      0    &     0     &  6\sqrt{3}& -3\sqrt{15}\cr
    &             &     -7    & -10\sqrt{2}& -4\sqrt{2}& -4\sqrt{10}\cr
    &             &           &     10    &     4     &  4\sqrt{5} \cr
    &             &           &           &    -2     & -2\sqrt{5} \cr
    &             &           &           &           &     17     \cr
    }\right]
   \label{eq:DsFromSK}
\end{equation}
in the basis
$[ \; (\Sigma K)_{P {1\over2}}, (\Sigma^* K)_{P {3\over2}}, 
   (\Sigma K^*)_{P {1\over2}}, (\Sigma K^*)_{P {3\over2}}, 
   (\Sigma^* K^*)_{P {1\over2}}, (\Sigma^* K^*)_{P {3\over2}} \;]$.
The corresponding relative couplings to the proton are
$\left[-{1\over3}, -\sqrt{8\over9}, \sqrt{25\over27} ,
\sqrt{8\over27}, \sqrt{8\over27}, \sqrt{40\over27}  \right]$.

Again, the net $\Delta s$ from these states is zero in the closure
limit, but this time the insertion of energy denominators does not
spoil the cancellation very much: the full calculation gives
$\Delta s =  -0.003$ in this sector.

   $P$-wave hyperons and kaons contribute another $-0.04$ to
$\Delta s$, and the net contribution from all higher states is
$-0.025$. {\it Thus, the result of our calculation is 
$\Delta s = -0.13$}, in quite good agreement with the most
recent extractions from experiment, 
$\Delta s\,({\rm expt}) = -0.10\pm 0.03$ (see, {\it e.g.}, 
Ref.~\cite{EllKar}). 
   We emphasize that our parameters were fixed
by spectra and decay data. Moreover, the result
is quite stable to parameter changes, varying by at most $\pm 0.025$
when $r_q$, $b$, $\beta_{\rm baryon}$ and $\beta_{\rm meson}$
are individually varied by 30\%. 

   For comparison with other calculations, we note that in our
model the $\Lambda K$ intermediate state alone contributes 
$-0.030$ to $\Delta s$, and the contribution from the
$\Lambda K$, $\Sigma K$, and $\Sigma^* K$ states together 
is (coincidentally) also $-0.030$.

   It is interesting to note that $\Delta s$ is driven 
mainly by meson, rather than baryon mass splittings: if we set
$m_{\Lambda} = m_{\Sigma} = m_{\Sigma^*}$, we find that $\Delta s$
decreases by only about 30\%, whereas it drops by about 80\%
if we set $m_K = m_{K^*}$. Finally, we have also  
computed the charm-quark contribution to the proton spin,
finding $\Delta c \approx -0.01$.

%%%%%%%%%%%%%%%%%%%%%%%%%%%%%%%%%%%%%%%%%%%%%%%%%%%%%%%%%%%%%%%%%%%%%
\subsection{Strangeness radius}
%%%%%%%%%%%%%%%%%%%%%%%%%%%%%%%%%%%%%%%%%%%%%%%%%%%%%%%%%%%%%%%%%%%%%

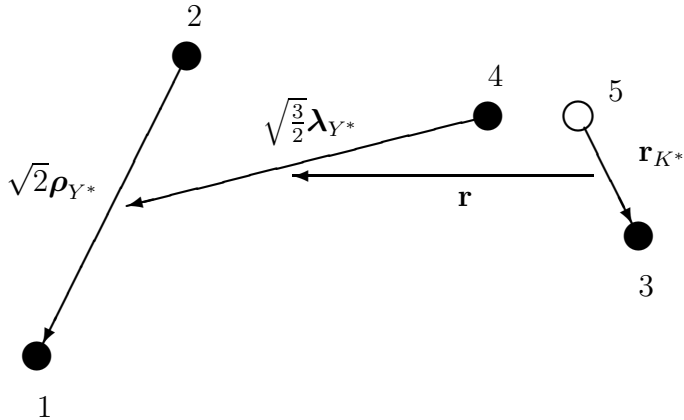
\begin{figure}
  \begin{centering}
  \setlength{\unitlength}{0.4 mm}
  \begin{picture}(300,200)(0,0)
    \thicklines
    \put( 30, 60){\circle*{10}}  \put( 80,160){\circle*{10}} 
    \put(180,140){\circle*{10}}  \put(230,100){\circle*{10}}  
    \put(210,140){\circle{10}}
    \put( 80,160){\vector(-1,-2){48}} 
    \put(180,140){\vector(-4,-1){120}}
    \put(212,136){\vector( 1,-2){16}}
    \put(215,120){\vector(-1, 0){100}}    

    \put( 30, 40){$1$}    \put( 80,170){$2$}
    \put(180,150){$4$}    \put(220,145){$5$}
    \put(230, 80){$3$}
  
    \put( 20,115){$\sqrt{2}\mbox{\boldmath$\rho$}_{Y^*}$}
    \put(105,135){$\sqrt{3\over2}\mbox{\boldmath$\lambda$}_{Y^*}$}
    \put(230,125){${\bf r}_{K^*}$}
    \put(170,110){${\bf r}$}

  \end{picture}
  \caption[x]{Quark coordinates in an intermediate state.}
  \label{fig:Rs2}
  \end{centering}
\end{figure}

   Figure \ref{fig:Rs2} defines our variables for the quarks in
an intermediate state. The (squared) distances of the $s$ and 
$\bar{s}$ quarks from the baryon-meson center of mass are 
\begin{eqnarray}
  r_s^2 = ({\bf r}_4 -{\bf R}_{cm})^2
      &=& \left[
     -\sqrt{6} \left({m_u\over m_{uus}}\right) 
     \mbox{\boldmath $\lambda$}_{Y^*}
     +\epsilon_{K^*} {\bf r}
     \right]^2 \\
     \nonumber \\
  r_{\bar{s}}^2 = ({\bf r}_5 -{\bf R}_{cm})^2
      &=& \left[
     -\left({m_u\over m_{us}}\right) {\bf r}_{K^*}
     -\epsilon_{Y^*} {\bf r}
     \right]^2 \; ,
  \label{eq:rsrsbar}
\end{eqnarray}
where $\epsilon_{K^*} \equiv M_{K^*}/M_{Y^*K^*}$ and
$\epsilon_{Y^*} \equiv M_{Y^*}/M_{Y^*K^*}$,
while by definition
\begin{equation}
  R_s^2 \equiv  r_s^2 - r_{\bar{s}}^2 \; 
  \label{eq:Rs}
\end{equation}   
\noindent is the strangeness radius.

   The calculation of $R_s^2$ is more difficult than the calculation
of $\Delta s$, for several reasons. First, the operators appearing
in $R^2_s$ cause orbital and radial transitions among the
intermediate states. Thus SHO transitions satisfying 
$\Delta n = 0,\pm 1$ and/or $\Delta \ell = 0,\pm 1$ are allowed,
so there are many more terms to calculate 
($n$ and $\ell$ are orbital and radial SHO quantum numbers).
  Moreover, the 
sub-cancellations discussed above no longer occur, so that
$R^2_s$ converges more slowly than $\Delta s$: we must include
more states in Eq.(\ref{eq:Os}) to obtain good accuracy.
   In addition, the basic matrix elements are more complicated:
in a basis of states with good magnetic quantum numbers 
$(m, m^\prime)$ we have, for example,
\begin{eqnarray}
  &&\langle n^{\prime} \ell^{\prime} m^{\prime} | r_{K^*z} 
  | n \ell m \rangle = \nonumber\\
  && \delta_{\ell^{\prime}\;\ell-1} \delta_{m^{\prime}\;m}
  \beta_{K^*}\sqrt{ {(\ell+m)(\ell-m) \over (2\ell+1)(2\ell-1)} }
  \left( \sqrt{n + \ell + 1/2} \delta_{n^{\prime}\;n} 
        -\sqrt{n + 1} \delta_{n^{\prime}\;n+1} \right) \nonumber\\
  &&+  \delta_{\ell^{\prime}\;\ell+1} \delta_{m^{\prime}\;m}
  \beta_{K^*}\sqrt{ {(\ell + m + 1)(\ell - m + 1) 
   \over (2\ell+1)(2\ell+3)} }
  \left( \sqrt{n + \ell + 3/2} \delta_{n^{\prime}\;n} 
        -\sqrt{n} \delta_{n^{\prime}\;n-1} \right)
  \label{eq:nlmr}
\end{eqnarray}
\vskip 0.5cm
\noindent for matrix elements of the meson internal coordinate and
\vskip 1.0cm
\begin{eqnarray}
  \langle q^{\prime} \ell^{\prime} m^{\prime} | r_z | q \ell m \rangle
  &=& i \delta_{m^{\prime}\;m}
  \left\{
  \delta_{\ell^{\prime}\;\ell-1}
  \sqrt{ {(\ell+m)(\ell-m) \over (2\ell+1)(2\ell-1)} }
  \left[ -{d\over dq} +{\ell-1 \over q} \right]\right. \nonumber\\
  &-&
  \left. \delta_{\ell^{\prime}\;\ell+1}
  \sqrt{ {(\ell+m+1)(\ell-m+1) \over (2\ell+1)(2\ell+3)} }
  \left[ {d\over dq} +{\ell+2 \over q} \right]
  \right\} 
  {\delta(q-q^{\prime}) \over q^2} 
  \label{eq:qlmr}
\end{eqnarray}
\vskip 0.5cm
\noindent for matrix elements of the $Y^*-K^*$ relative coordinate.
  These matrix elements must be coupled together to give
$\langle R_s^2 \rangle$ between states of definite 
$\ell$ and $S$ with total angular momentum $1 \over 2$, 
leading to formulas which
become quite lengthy, especially for 
excited intermediate states. 
Thus we were happy to have a stringent check
of our results: when we equate all of the energy denominators
in Eq.~(\ref{eq:Os}), we must obtain the closure-spectator
result, $R^2_s=0$.

   Our results for $R^2_s$ are shown in Table~\ref{tab:Rs}.
With our standard parameter set, we obtain 
$R_s^2 = -0.04 {\rm fm}^2$.
   For reasonable parameter variations, $R_s^2$ ranges between
$-0.02$ and $-0.06 {\rm fm}^2$.
  Table~\ref{tab:Rs} shows that the lowest lying $SU(6)$ 
multiplets of intermediate states ({\it i.e.}, the $S$-wave
hyperons and kaons) account for about half of $r_s^2$ and 
$r_{\bar{s}}^2$.
   Most of the remaining contributions come from $P$-wave hyperons
and kaons. However, $R_s^2$ involves a large cancellation
between $r_s^2$ and $r_{\bar{s}}^2$, and its value doesn't settle
down until we add in quite highly excited intermediate states.
   For this reason, the precise numerical value (and perhaps even the sign) of 
$R_s^2$ cannot be considered definitive: our conclusion 
is rather that $R_s^2$ is small, 
about an order of magnitude smaller than $r_s^2$ and $r_{\bar{s}}^2$. 
   This result is not too surprising: $R_s^2$ is exactly 
zero in the closure limit, and 
our previous hadronic loop studies~\cite{GIpotential,GIonOZI} led us 
to expect that the full calculation with energy denominators 
would preserve the qualitative features of this limit.

Note that the $\Lambda K$ intermediate state
alone gives $R_s^2 \approx -0.01 {\rm fm}^2$  (the sign is  
as expected from the usual folklore) while
the $\Lambda K$, $\Sigma K$, and $\Sigma^* K$ states together 
give $-0.017 {\rm fm}^2$. Nevertheless,
although our sum over all states gives the same sign and order of
magnitude as these truncations, Table~\ref{tab:Rs} shows that this
is just a coincidence.

\vspace{2.5cm}
\begin{table}[th]
  \caption[x]{Proton strangeness radius from hadronic loops 
  (in fm$^2$).
   The rows give the running totals as progressively more excited
   intermediate states are added into the calculation. The final
   column thus shows the total from all intermediate states. }
  \begin{tabular}{lcccc}
  & S-waves & plus           & plus D-waves and            & all    \\
  &         & P-waves        & S-wave radial excitations   & states \\
  \hline
  $r_s^2$          &.097 &.198 &.210 & .173 \\
  $r_{\bar{s}}^2$  &.094 &.139 &.185 & .210 \\
  $R_s^2$          &.003 &.059 &.025 & -.04 \\
  \end{tabular}
  \label{tab:Rs}
\end{table}
% 

%%%%%%%%%%%%%%%%%%%%%%%%%%%%%%%%%%%%%%%%%%%%%%%%%%%%%%%%%%%%%%%%%%%%%
\subsection{Strange magnetic moment}
%%%%%%%%%%%%%%%%%%%%%%%%%%%%%%%%%%%%%%%%%%%%%%%%%%%%%%%%%%%%%%%%%%%%%

   The strange and antistrange quarks carry magnetic 
moments $-{1 \over 3} \mu ^{(s,\bar s)}$ where
\begin{equation}
  \mu^{(s)} = {1\over 2m_s}\langle 2S_z^{(s)} + L_z^{(s)} \rangle
  \label{eq:Mus}
\end{equation}
\begin{equation}
  \mu^{(\bar{s})} = -~~{1\over 2m_s}\langle 2S_z^{(\bar{s})} 
  + L_z^{(\bar{s})} \rangle
  \label{eq:Musbar}
\end{equation}
and we denote the net strange magnetic moment by $\mu_s$:
\begin{equation}
  \mu_s \equiv \mu^{(s)} + \mu^{(\bar{s})} .
  \label{eq:MusTot}
\end{equation}
The spin expectation values are already in hand from our
$\Delta s$ calculation.
  Referring again to Fig.~\ref{fig:Rs2}, we see that the $s$ 
and $\bar{s}$
orbital angular momenta are given by
\begin{eqnarray}
  \mbox{\boldmath $L$}_s &=& 
  ({\bf r}_4 -{\bf R}_{cm})\times {\bf p}_4
      \nonumber \\
      &=& \left[
     -\sqrt{6} \left({m_u\over m_{uus}}\right) 
     \mbox{\boldmath $\lambda$}_{Y^*}
     +\epsilon_{K^*} {\bf r} \right]
     \times
     \left[-{2\over\sqrt{6}}{\bf p}_{\lambda_{Y^*}}
     + \left({m_s\over m_{uus}}\right){\bf q} \right] \\
     \nonumber \\
  \mbox{\boldmath $L$}_{\bar s} &=& 
  ({\bf r}_5 -{\bf R}_{cm})\times {\bf p}_5
      \nonumber \\
      &=& \left[
     -\left({m_u\over m_{us}}\right) {\bf r}_{K^*}
     -\epsilon_{Y^*} {\bf r} \right]
     \times
     \left[-{\bf p}_{K^*}
     -\left({m_s\over m_{us}}\right){\bf q} \right] \; .
  \label{eq:LsLsbar}
\end{eqnarray}
Computing the expectation values of these operators presents
no new difficulties beyond those encountered in the $R_s^2$
calculation. In fact, there are no radial transitions in this 
case, so there are fewer states to sum over and the sum 
converges more quickly.

   The results obtained with our standard parameter set 
are 
\begin{equation}
  \begin{array}{lcl}
  \langle 2S_z^{(s)} \rangle          = -0.058 && 
  \langle 2S_z^{(\bar{s})} \rangle    = -0.074 \\
  \langle L_z^{(s)} \rangle       =  0.043 && 
  \langle L_z^{(\bar{s})} \rangle =  0.038 \\
  \mu^{(s)} = -0.025\mu_N  &&  \mu^{(\bar{s})} = 0.060\mu_N \\
  & \mu_s = 0.035\mu_N &
  \end{array}
  \label{eq:MusResults}
\end{equation}
 We predict a positive (albeit small) value for $\mu_s$,
in disagreement with the other models discussed at the beginning 
of this section. 
   Where does the positive sign originate?
   First note that the signs of $\langle S_z^{(s)}\rangle$,
$\langle L_z^{(s)}\rangle$, and $\langle L_z^{(\bar{s})}\rangle$ 
are correctly given by just the lowest lying intermediate state, 
$\Lambda K$ of Eq.~(\ref{eq:LamK}). (We also note in passing that the 
$L_z$'s have similar magnitudes so that orbital angular momentum 
contributes very little to $\mu_s$ in any case.) 
   On the other hand, the $\Lambda K$ state
has $\langle S_z^{(\bar{s})} \rangle=0$, whereas we find 
$\langle S_z^{(\bar{s})} \rangle$ to be quite large and negative.
(The main contribution comes from the off-diagonal process
$p \rightarrow (\Lambda K)_{P {1\over2}} {\buildrel S_z^{(\bar{s})} \over \rightarrow}
(\Lambda K^*)_{P {3\over2}} \rightarrow p$, although 
there is also a significant contribution from
$p \rightarrow (\Lambda(1405) K)_{S {1\over2}} {\buildrel S_z^{(\bar{s})} \over \rightarrow}
(\Lambda(1405) K^*)_{S {1\over2}} \rightarrow p$.)
  These important terms, which drive $\mu_s$ positive, 
are omitted in calculations which include only kaon loops.
  (We find that the $\Lambda K$ intermediate state alone 
contributes $-0.080 \mu_N$ to $\mu_s$, and the 
contribution from  $\Lambda K$, $\Sigma K$, and $\Sigma^* K$ 
together is $-0.074 \mu_N$.)

%%%%%%%%%%%%%%%%%%%%%%%%%%%%%%%%%%%%%%%%%%%%%%%%%%%%%%%%%%%%%%%%%%
\section{Conclusions} 
%%%%%%%%%%%%%%%%%%%%%%%%%%%%%%%%%%%%%%%%%%%%%%%%%%%%%%%%%%%%%%%%%%
\label{sec:Concl}

     We have presented here parameter-free calculations
of the effects of the $s \bar s$
sea generated by strong 
 $Y^*K^*$ loops on the low energy, nonperturbative structure of the nucleons.
These calculations represent what
is to our knowledge the first such results within a framework which
has been demonstrated to be consistent with the many empirical
constraints which should be applied to such calculations, namely
consistency with the success of the quark potential model and 
especially with the validity of the OZI rule.

    Our results predict that observable effects from the strange
sea generated by such loops  arise from delicate cancellations between large contributions
involving a suprisingly massive tower of virtual meson-baryon
intermediate states. If correct, our conclusions rule out the
utility of a search for a simple but predictive low energy hadronic truncation 
of QCD. While complete (in the sense of summing over all OZI-allowed 
$Y^*K^*$ loops) and gauge invariant, we recall that our calculation has ignored 
pure OZI-forbidden effects as well as those loop diagrams directly generated by the
probing current (contact terms). As
a consequence, our results cannot strictly speaking be taken as predictions for 
$\Delta s$, $R^2_s$, or $\mu_s$. Rather, this calculation shows that
a complete set of strong $Y^*K^*$ loops, computed in a model consistent
with the OZI rule, gives very small observable $s \bar s$ effects.
While such OZI-allowed processes {\it might}  dominate,   we cannot
rule out the possibility (as was also the case with 
$\omega-\phi$ and other meson mixing~\cite{GIonOZI})
that direct OZI violation (and in this case contact graphs
as well) could make additional contributions of a comparable magnitude.

   The small residual effect we calculate for the loop contributions to $\Delta s$ seems 
consistent
with the most recent analyses of polarized deep inelastic scattering 
data. Our calculations also give small residual strange quark
contributions to the charge and magnetization distributions
inside the nucleons. If these contributions are dominant, it will
be a challenge to devise experiments that are capable of seeing
them. Indeed, they are sufficiently small that we would expect
that their observation will require the development of special
apparatus dedicated to this task. Given the fundamental nature
of the puzzling absence of other signals for the strong $q \bar q$
sea in   low energy phenomena, this effort seems very
worthwhile.

   It would  be desirable
to devise tests of the mechanisms underlying the delicate 
cancellations which 
conspire to hide the effects of the sea in  the picture
presented here. It also seems very worthwhile to extend 
this calculation
to $u \bar u$ and $d \bar d$ loops. 
Such an extension could reveal
the origin of the observed violations~\cite{GSRexpt}
of the Gottfried Sum Rule~\cite{GSR} and also complete 
our understanding of
the origin of the spin crisis.
From our previous calculations~\cite{GIpotential}, the effects of
``unquenching" strange quarks are
a good guide to the effects to be
expected from
up and down quarks in the absence of Pauli blocking. Since most of the 
created pairs are in highly excited states,
Pauli blocking should be of minor importance, and so one would guess that
{\it each} of up, down, and strange will produce a contribution to
$\Delta q$ of about
$-0.13$. When combined with the relativistic quenching of
$\Delta q_{valence}$~\cite{gA},  this makes it plausible to us 
that most of the ``missing spin" of the proton is in 
orbital
angular momentum.

\bigskip\bigskip

\acknowledgements
 
P.G. thanks Martin Savage for discussions, as well as 
NSERC of Canada and the U.S. Dept. of Energy under grant No.
DE-FG02-91ER40682 for financial support. N.I. gratefully acknowledges
several helpful discussions with Michael Musolf,
especially one which pointed out a 
potential flaw in an unpublished version
of this paper. 

\hfill
\eject

{\centerline {\bf APPENDIX: DERIVATION OF $\langle O_s \rangle$}}

   While Eq. (5) has a simple nonrelativistic interpretation as an expectation 
value of $O_s$ in the dressed state of Eq. (4), is it not obvious that it 
actually computes $\Delta s$, $R^2_s$, and $\mu_s$.  In particular, $R^2_s$ 
and $\mu_s$ are defined in terms of charge and magnetic form 
factors, respectively. Here we show that the contributions 
to all three quantities from $s \bar s$ pairs arising from
strong $Y^*K^*$ loops
may indeed be 
calculated in a {\it gauge invariant} fashion in our model {\it via} this simple formula,
with each given by the usual nonrelativistic operators of Eqs. (12), (18), and (23).
	
   We first discuss the relatively simple case of $\Delta s$, where gauge 
invariance plays no role. To second order in $h_{s \bar s}$, 
the graphs 
contributing to the nucleon matrix elements of 
$A^{\mu}_s \equiv \bar s \gamma^{\mu} \gamma_5 s$ are of two types: 1) OZI-allowed graphs
where a strong $s \bar s$ loop has created a 
$Y^*K^*$
loop and in which $A^{\mu}_s$ scatters the $s$ or $\bar s$ quark, and 2) pure
OZI-forbidden graphs in which $A^{\mu}_s$ acts on a color singlet $s \bar s$ state
which is created and/or destroyed on the nucleon. The pure
OZI-forbidden processes fall outside the scope of our model. On the other hand, OZI-allowed
contributions may be calculated
(with all the usual {\it 
caveats} and approximations of nonrelativistic 
quark model calculations of axial matrix
elements) according to

\begin{equation}
~~~~~~~~~~~~~~~~~~~~~g^{(s)}_A \equiv \langle p({\bf 0},+) \vert \bar s \gamma^{\mu} \gamma_5 s 
\vert p({\bf 0},+) \rangle
\end{equation}

\begin{equation}
~~~~~~~~~~~~~~~~~~~~~~~~~~~~~~~~\simeq 2 \langle p({\bf 0},+) \vert S_z^{(s)} + S_z^{(\bar s)}
\vert p({\bf 0},+) \rangle
\end{equation}

\begin{equation}
=\Delta s
\end{equation}
as used in Section V.A.

	The discussion of $R^2_s$ and $\mu_s$  is considerably more involved. 
Our first step is to set up the Breit-frame formalism for calculating
$R^2_s$ and $\mu_s$. It is easily shown that the electric and magnetic form factors
of the proton may be calculated {\it via} the formulas
\begin{equation}
G^p_E (q^2=-{\bf Q}^2) =\frac {T_E^{ext}({\bf Q})}{e \phi_E^{ext}({\bf Q})}
\end{equation}
\begin{equation}
~~~~~~~G^p_M (q^2=-{\bf Q}^2)=- \frac {(\frac {2M_p}{Q}) T_M^{ext}({\bf Q})}
{e a_M^{ext}({\bf Q})}
\end{equation}
where $T_{E,M}^{ext}({\bf Q})$ are the $T$-matrix elements
\begin{equation}
T_E^{ext}({\bf Q}) \equiv \langle p(+ \frac{Q \hat {\bf z}}{2}, +) \vert {\bf T}_E
\vert p(- \frac{Q \hat {\bf z}}{2}, +) \rangle
\end{equation}
\begin{equation}
T_M^{ext}({\bf Q}) \equiv \langle p(+ \frac{Q \hat {\bf z}}{2}, -) \vert {\bf T}_M
\vert p(- \frac{Q \hat {\bf z}}{2}, +) \rangle
\end{equation}
for scattering in the external potentials
\begin{equation}
\phi_E^{ext}(t, {\bf x})= \int d^3Q e^{i {\bf Q} \cdot {\bf x}} \phi_E^{ext}({\bf Q})
\end{equation}
and
\begin{equation}
{\bf A}_M^{ext}(t, {\bf x})= \frac {1}{2}( \hat {\bf x} -i \hat {\bf y})
\int d^3Q e^{i {\bf Q} \cdot {\bf x}} a_M^{ext}({\bf Q})
\end{equation}
respectively.  (Note that all normalizations are fixed
by the condition that $G_E^p(0)=1$.) 

	    Now consider a proton scattering off one of these external potentials subject
to the additional gauge invariant perturbation  created by $s \bar s$  pair 
creation.  The resulting changes $\Delta G_E^p$ and $\Delta G_M^p$ 
are $G_E^{(s)}$ and $G_M^{(s)}$, which may 
therefore be associated with $\Delta T_E^{ext}$ and $\Delta T_M^{ext}$. 
Thus, generically, the $\Delta G$'s may be 
associated with four processes.  Associated with the ``naive" formula (5) are the two processes
in which:

\bigskip
  1a) the external field probes the $K^*$ of the $Y^*K^*$ loop, and

  1b) the external field probes the $Y^*$ of the $Y^*K^*$ loop.
\bigskip

\noindent We will recall at the end of this Appendix how these processes, which are
of second order in the strong $s \bar s$ pair creation Hamiltonian density
$h_{s \bar s}$ and third order in perturbation theory, lead to Eq. (5).

   Associated with $h_{s \bar s}$, which has matrix elements proportional to
$(-1)^{2 \bar s}\chi^{\dagger}_s {\bf \sigma} \chi_{-\bar s}\cdot ({\bf p_s}-{\bf p_{\bar s}})$,
is a pair creation contact interaction  density $h_{s \bar s}^{contact}$
with matrix elements proportional to
$2e_s(-1)^{2 \bar s}\chi^{\dagger}_s {\bf \sigma} \chi_{-\bar s}\cdot{\bf A}
\equiv {\bf j}_{s \bar s}^{contact} \cdot {\bf A}$,
where $e_s=-\frac{1}{3} e$ and $\bf A$ is the vector potential. Through this contact
interaction two more processes can occur:

\bigskip
   2a) the external field directly creates the $s \bar s$ pair through $h_{s \bar s}^{contact}$,
converting the proton at momentum $-\frac{Q \hat {\bf z}}{2}$ into a $Y^*K^*$ system
at momentum $+\frac{Q \hat {\bf z}}{2}$, and then $h_{s \bar s}$ annihilates the
$s \bar s$ pair, returning the $Y^*K^*$ system to a proton at the same momentum
$+\frac{Q \hat {\bf z}}{2}$, and

   2b) $h_{s \bar s}$ 
converts the proton at momentum $-\frac{Q \hat {\bf z}}{2}$ into a $Y^*K^*$ system
at the same momentum, and then $h_{s \bar s}^{contact}$
 annihilates the
$s \bar s$ pair, turning the $Y^*K^*$ system at momentum $-\frac{Q \hat {\bf z}}{2}$
to a proton at momentum
$+\frac{Q \hat {\bf z}}{2}$.
 
\bigskip

\noindent We begin by showing that the contact terms 2a) and 2b) are not required to make the
loop contributions 1a) and 1b) 
to $R_s^2$ or $\mu_s$ gauge invariant.

   The case of 
$R^2_s$ is trivial:  $h_{s \bar s}^{contact}$ is independent of
$\phi_E^{ext}(t, {\bf x})$ in our model, so contact processes do not 
contribute to it.

  In contrast, $\mu_s$ is determined by scattering in
${\bf A}_M^{ext}(t, {\bf x})$, which does couple to $h_{s \bar s}^{contact}$.
From ordinary second-order
time-ordered perturbation theory
\begin{equation}
\frac {2T_M^{ext}({\bf Q})}{e a^{ext}_M({\bf Q})} =  \int dq q^2 \sum_{\ell S} 
\frac{1}{\Delta(Q^2)}
\langle p(+ \frac{Q \hat {\bf z}}{2}, -) \vert \left[ t_{sc}^{q \ell S}+t_{cs}^{q \ell S}   \right]
\vert p(- \frac{Q \hat {\bf z}}{2}, +) \rangle
\end{equation}
where $\Delta(Q^2)$ is an energy denominator which can easily 
be shown to be even in $Q$, and where 
\begin{equation}
t_{sc}^{q \ell S}= h_{s \bar s}(0, {\bf 0}) 
\vert [Y^*K^*]_{q\ell S}(+\frac{Q \hat {\bf z}}{2}, -)
\rangle\langle  [Y^*K^*]_{q\ell S}(+\frac{Q \hat {\bf z}}{2}, -)
 \vert {\bf j}^{contact}_{s \bar s}(0, {\bf 0})_{-}
\end{equation}
and
\begin{equation}
t_{cs}^{q \ell S}= {\bf j}^{contact}_{s \bar s}(0, {\bf 0})_{-} 
 \vert [Y^*K^*]_{q\ell S}(-\frac{Q \hat {\bf z}}{2}, +)
\rangle\langle  [Y^*K^*]_{q\ell S}(-\frac{Q \hat {\bf z}}{2}, +)
 \vert h_{s \bar s}(0, {\bf 0})~,
\end{equation}
where $\vert [Y^*K^*]_{q\ell S}(\pm\frac{Q \hat {\bf z}}{2}, m_{Y^*K^*}) \rangle$ 
denotes a state with internal radial momentum $q$, internal orbital angular momentum
$\ell$ and total internal spin $S$ coupled to total angular momentum and parity 
$J^P=\frac{1}{2} ^+$ (so that $h_{s \bar s}$ can connect it to the proton),
with momentum $\pm\frac{Q \hat {\bf z}}{2}$,
and with $z$-component of spin $m_{Y^*K^*}$.
    Now consider the general matrix element
\begin{equation}
\langle [Y^*K^*]_{q\ell S}(+\frac{Q \hat {\bf z}}{2}, m_{Y^*K^*})
\vert h_{s \bar s}(0,{\bf 0})
\vert p(-\frac{Q \hat {\bf z}}{2},m_p) \rangle
= g^{q\ell S}_{Y^*K^*p}(Q^2)\chi^{\dagger}_{Y^*K^*}\chi_p~.
\end{equation}
Since $h_{s \bar s}$ is a scalar operator,
the form of the right-hand side of this equation is uniquely determined. In 
particular, it follows from parity conservation that this matrix element has no spin-flip 
component and therefore that when made gauge invariant {\it via} contact interactions, it
will not generate one. We conclude that also no contributions to $\mu_s$ from
contact terms are required to make the contributions of
the strong $Y^*K^*$ loops gauge invariant.

   There is, of course, no reason why the underlying $s \bar s$ pair creation dynamics
might not generate independent gauge invariant contact currents of the form
$\chi^{\dagger}_{Y^*K^*}\epsilon^{ijk}\sigma^jQ^k\chi_p$
which {\it could} contribute to $\mu_s$. Although they
lie outside our goal of providing a complete calculation of the effects of $s \bar s$
pairs generated by strong $Y^*K^*$
loops, it would be interesting to
know the size of such effects. (We note in passing that, as a class of diagrams,
they will also be subject to strong
cancellations since, for example,  $^3S_1$ pair creation followed by
$^3P_0$ annihilation will vanish in the closure-spectator approximation). 

    To complete our demonstration that Eq. (5) is the correct gauge invariant
formula for the contributions of $s \bar s$ pairs from strong $Y^*K^*$ loops to
$\Delta s$, $R_s^2$, and $\mu_s$, we 
briefly recall how the third-order processes 1a) and 1b)
lead to it.  The analysis is very straightforward,
giving
\begin{equation}
T_M^{ext}({\bf Q})_{1a)+1b)} =T_M^{ext}({\bf Q})_{Y^*} + T_M^{ext}({\bf Q})_{K^*}
\end{equation}
corresponding to scattering from intermediate particles $Y^*$ and $K^*$, 
respectively, with, for example,
\begin{equation}
T_M^{ext}({\bf Q})_{K^*} =e j^{K^*}_M(Q)a_M^{ext}({\bf Q}) \sum_{m_{Y^*}m_{K^*}}
\int d^3\pi \Phi^{\ell S~~*}_{m_{Y^*}m_{K^*}, -}({\bf \pi}) \Phi^{\ell S}_{m_{Y^*}m_{K^*}, +}
({\bf \pi}+ \epsilon_{Y^*} {\bf Q})
\end{equation}
\begin{equation}
 =e j^{K^*}_M(Q)a_M^{ext}({\bf Q}) \sum_{m_{Y^*}m_{K^*}}
\int d^3r \psi^{\ell S~~*}_{m_{Y^*}m_{K^*}, -}({\bf r}) 
e^{-i {\bf Q} \cdot {\bf r}_{K^*}}\psi^{\ell S}_{m_{Y^*}m_{K^*}, +}({\bf r})
\end{equation}
where  ${\bf r}_{K^*}=\epsilon_{Y^*} {\bf r}$ and where
\begin{equation}
\Phi^{\ell S}_{m_{Y^*}m_{K^*}, s}({\bf \pi})=\langle \bigl[Y^*( \epsilon_{Y^*}
 {\bf P} + {\bf \pi}, m_{Y^*}) 
K^*( \epsilon_{K^*} {\bf P} - {\bf \pi}, m_{K^*}) \bigr]_{\ell S} 
\vert h_{s \bar s}(0,{\bf 0})
\vert p({\bf P},s) \rangle 
\end{equation}
is the internal relative momentum wavefunction of $Y^*$ and $K^*$ ``inside" the 
proton (which is 
 independent of the  (small) total momentum ${\bf P}$), and $\psi$ is its 
Fourier transform.  Clearly an exactly analogous formula appears 
for $T_M^{ext}({\bf Q})_{Y^*}$ and the 
two, having 	as they do simple constituent interpretations, lead to the 
``naive" Eq. (5).

\hfill
\eject

{
\hfill
\eject


\begin{references}
\bibitem{LevelCrossing}
Unlike the Coulombic region where the $q\bar q$ 
pairs are all virtual, in the long-distance region pair
production can also lead to real dissociation into a 
$Q\bar q$ plus $q\bar Q$ 
system if the energy $E_0(r)$ corresponds exactly to the 
sum of the masses of states of these  two meson 
subsystems.  In the language of adiabatic surfaces, these  
disocciations occur when the $Q\bar Q$ adiabatic 
surface unperturbed by $q\bar q$ 
pair creation crosses the energies 
$E_{ij}=m^{Q\bar q}_i+m^{q\bar Q}_j$ 
corresponding to the two-meson adiabatic surfaces (which 
are independent of $r$ for large $r$ and have no continuum parts since 
the $Q$ and $\bar Q$ are fixed).  The shifted adiabatic 
surface can, of course, be tracked through the level 
crossings to follow the quantum state that remains 
dominantly a $Q\bar Q$ pair connected by a flux tube.
\end{references}
\end{document}